\documentclass{aa}
\usepackage{epsf}
\begin{document}

   \thesaurus{01 (13.07.1; 
                  09.07.1  
                 )} 
   \title{Optical counterparts of cosmological GRBs due to
          heating of ISM in the parent galaxy}

   \author{G.S. Bisnovatyi-Kogan\inst{1}
           \and
           A.N. Timokhin
           \inst{1}\inst{2}
          }

   \offprints{A.N.Timokhin}

   \institute{Space Research Institute, 
              Profsoyuznaya 84/32,
              117810 Moscow, Russia
              email: gkogan@mx.iki.rssi.ru
         \and
              Sternberg Astronomical Institute,
              Universitetskij pr. 13,
              119899 Moscow, Russia  
              email: timokhin@mx.iki.rssi.ru
             }

   \date{ Received date; accepted date }
   \authorrunning{G.S. Bisnovatyi-Kogan and  A.N. Timokhin}
   \titlerunning{Optical counterparts of cosmological GRBs}
   \maketitle

   \begin{abstract}
We investigated influence of cosmological GRB on the surrounding
interstellar medium. It was shown that $\gamma$-radiation from the 
burst heats interstellar gas to the temperatures $>10^4$K
 up to the distance $\sim10$ pc. For high
density ISM optical and UV radiation of the heated gas can be 
observed on the Eath several years as a GRB`s counterpart.
   \end{abstract}

%

\section{Introduction}

In the cosmological model for Gamma-Ray Bursts energy production 
in $\gamma$-region must be
enormous large $10^{51} - 10^{53}$ ergs, during several seconds. Such
huge energy flux interacts with the matter in the parent  galaxy 
leading to its
heating and formation of counterparts in another spectral region. We
investigate results of this interaction without specifying the
mechanism of GRB formation.

The problem is solved of the respond of interstellar medium with normal
chemical composition on propagation through it of a short powerful
gamma-ray pulse. Spherical symmetry was accepted; different densities
of interstellar gas and  burst energies were considered.

\section{Basic equations}
The basic equations, describing this problem are :
\begin{equation}
\frac{\partial \varrho }{\partial t} + \frac{1}{r^2}
\frac{\partial}{\partial r} (r^2\varrho u) = 0 \ ,
\label{1}
\end{equation}

\begin{eqnarray}
\frac{\partial u}{\partial t} + u\frac{\partial u}{\partial r} =
- \frac{1}{\varrho }\frac{\partial p}{\partial r} + F_{\gamma} \ ,
\label{2}
\end{eqnarray}

\begin{eqnarray}
\frac{\partial \epsilon}{\partial t} + u\frac{\partial \epsilon}
{\partial r} = - \frac{p}{\varrho }\frac{1}{r^2}\frac{\partial}
{\partial r}(r^2u) + H_{\gamma} - C_{\gamma} \ .
\label{3}
\end{eqnarray}
Where 
\begin{eqnarray}
\begin{array}{c}
p = (n_i + n_n + n_e)kT,\qquad
\epsilon = \frac{3}{2} \frac{kT}{\mu m_u},\\
\end{array}
\end{eqnarray}
$n_n$, $n_i$, $n_e$ are concentrations of neutral atoms,
ions and electrons correspondently.

For the $\gamma$-ray signal interacting with matter
due to Thomson scattering of photons on electrons
with cross-section $\sigma _T$ we have:
\begin{equation}
F_{\gamma} =
\frac{1}{c}\frac{L}{4\pi r^2}\frac{\mu_e \sigma _T}{m_u}\ ,
\end{equation}

\noindent For $\gamma$-rays with $h\nu\gg\mbox{B}_e^{(\mbox{a,i})}$
(B${}_e$ is the binding energy of electrons in atoms or ions) the
cross-section is almost the same for free and bound electrons.
Heating of the gas by the light signal with the spectrum
{\normalsize
$
\frac{dL}{dE}=\frac{L}{E_{max}}e^{-E/E_{max}}
$}
is given by (Bisnovatyi-Kogan \& Blinnikov \cite{B-KB}):  
\begin{eqnarray}
H_{\gamma} = \frac{L}{4\pi r^2}\frac{\mu_e \sigma _T}{m_u}
\frac{E_{max} -4kT}{m_ec^2} \ ,
\label{heat}
\end{eqnarray}
where  $L$ is the energy flux of the signal; $E$ is the
energy of photons. This formula is valid for $E \ll m_ec^2$.
For photons with energies $<10$ keV the main process of interaction
with the gas is photoabsorption by ions of heavy elements.
To take in to account gas heating by soft X-rays photons  
we enhance term (\ref{heat}) using $E_{max} \sim 2$ MeV. 
Because of lack of information about soft X-ray spectra of GRBs
and variations in interstellar media density such assumption 
seems to be quite good. On the other hand, completely ignoring
gas heating by X-rays we  get a counterpart 
by less then $2^m$ fainter. 
Cooling of optically thin plasma by free-free  and free-bound
transitions is given by approximating function
(Cowie et al. \cite{CowieMcKee}):
\begin{eqnarray}
C_{\gamma} = \frac{\Lambda(T) n^2}{\varrho }\ , \qquad
n = n_n + n_i
\label{cool}
\end{eqnarray}

\begin{eqnarray*}
\Lambda(T) = \cases{%
                             0 ,&when $T < 10^4$K;\cr
1.0\cdot 10^{-24}\cdot T^{0.55},&when $10^4$K$< T< 10^5$K;\cr
6.2\cdot 10^{-19}\cdot T^{-0.6},&when $10^5$K$< T< 4\cdot 10^7$K;\cr
2.5\cdot 10^{-27}\cdot T^{0.5} ,&when $T > 4\cdot 10^7$K.\cr}
\label{lambda}
\end{eqnarray*}
Electron density ($n_e$) is given by the Elvert  formula.

This system was solved numericaly using a full 
conservative  difference scheme with
flux corrected transport, because there is a strong
density gradient in the solution (shock wave).

\begin{figure*}
 \centerline{ \epsfbox[0 0 520 220]{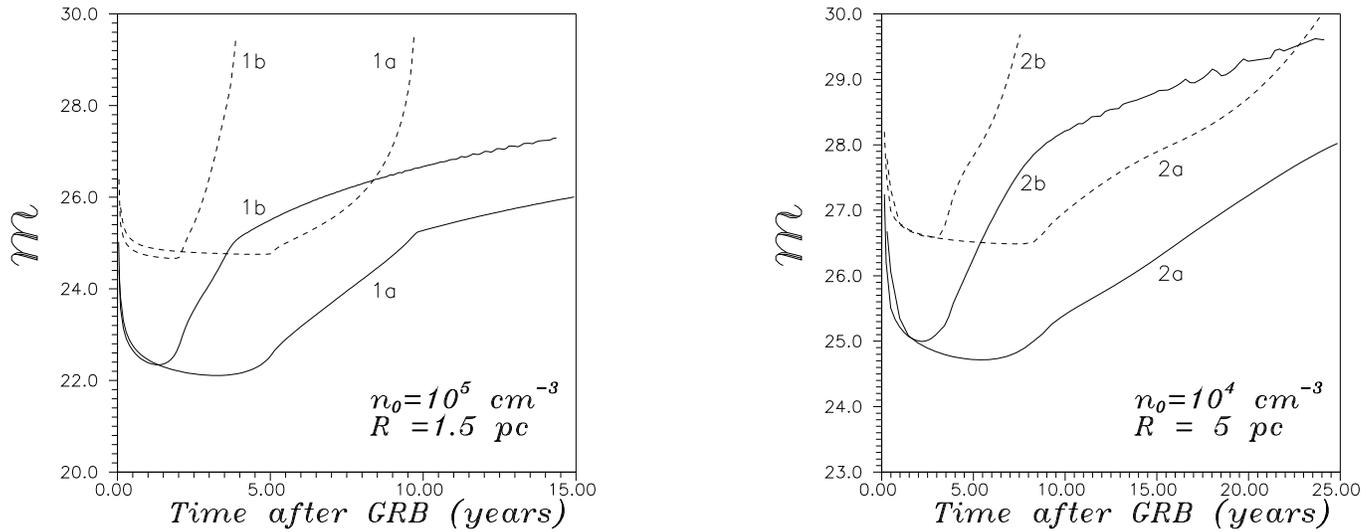} }
\caption{The magnitudes of the counterparts (upper limit - solid
line, lower limit - dashed line) as a function of time after burst for
GRB with total flux near the Earth
$F_{GRB}=10^{-4}\mbox{ erg cm}^{-2}$ :
1a - for the case $E=10^{52}\mbox{ erg},
 n_0 = 10^5\mbox{ cm}^{-3}$;
1b - for the case $E=10^{51}\mbox{ erg},
 n_0 = 10^5\mbox{ cm}^{-3}$;
2a - for the case $E=10^{52}\mbox{ erg},
 n_0 = 10^4\mbox{ cm}^{-3}$;
2b - for the case $E=10^{51}\mbox{ erg},
 n_0 = 10^4\mbox{ cm}^{-3}$.}
\label{f1}
\end{figure*}

\section{Main results}

For GRB with the energy $10^{52}$ erg we considered following
densities of the surrounding interstellar medium:
$10^5 \mbox{cm}^{-3},\ 300 \mbox{cm}^{-3},\ 0.25 \mbox{cm}^{-3},\
1.6\cdot 10^{-3} \mbox{cm}^{-3}$. Main results of calculations in
optically thin approximation are presented
in our publications 
Timokhin \& Bisnovatyi-Ko\-gan (\cite{Timokhin1,Timokhin2}).  According
these calculations a counterpart of cosmological GRB 
due to the influence of GRB on the surrounding 
medium could be observed if density
of the medium is greater then $300 \mbox{cm}^{-3}$. For such dense
interstellar clouds optically thin approximation is not valid because
of large optical depth in main emission lines. 

Then we considered
a simplified picture of conversion of UV photons into optical ones 
in opticaly thick regions allowed us to get optical and
UV light curves of counterparts (see details in 
Bisnovatyi-Kogan \& Timokhin \cite{Timokhin3}).  
We considered GRBs with energies $10^{52}$,$10^{51}$ ergs 
and following initial conditions in a gaseous cloud of radius $R$:
\begin{itemize}
\item[1]
$n_0 =10^5\mbox{cm}^{-3}, T_0 = 20\mbox{K},  R = 1.5 \mbox{pc}$

\item[2]
$n_0 =10^4\mbox{cm}^{-3}, T_0 = 20\mbox{K},  R = 5 \mbox{pc}$
\end{itemize}
The results of calculations are presented in Fig.~\ref{f1}.
Dashed lines give the lower boundary for luminosity, 
and solid lines - the upper one. For GRB
with total energy output $10^{51}$ and $10^{52}$ ergs maximal
luminosities of counterparts differ by one order of magnitude too. It
means that relative maximum brightness of the counterparts depends
weakly on the total energy output. The total duration of the
counterpart radiation decreases 2 times when the total output decreases
10 times.  A more detailed description of the results is given
in Bisnovatyi-Kogan \& Timokhin (\cite{Timokhin3}).

\section{Conclusions}

It is shown, that counterparts of cosmological GRB due to
interaction of gamma-radiation with dense interstellar media are
"longliving" objects, existing  for years after GRB. 
 To distinguish GRB counterpart from a superova
event, having similar energy output, it is necessary to take into
account its unusual light curve and spectrum. In optical region 
of the spectrum the strongest emission lines
are H$_{\alpha}$ and H$_{\beta}$.

The volume occupied by dense interstellar medium in the parent galaxy
is relatively small, so the probability to get bright counterpart of
such kind is also small. Nevertheless, discovery of even one
optical counterpart of GRB with properties described above could be a
decisive argument in favor of the cosmological nature of GRB.

\begin{acknowledgements}
Part of this work was
supported by Russian Foundation of Basic Research
(RFFI) under grant 96-02-16553.
\end{acknowledgements}

\end{document}